\documentstyle[12pt]{article}

\input epsf.sty

\newcommand{\MS}{\overline{\rm MS}}

\begin{document}
\begin{titlepage}

\centerline{\large \bf Predictions from conformal algebra for the}
\centerline{\large \bf deeply virtual Compton scattering.}

\vspace{5mm}

\centerline{\bf A.V. Belitsky}

\vspace{5mm}

\centerline{\it Bogoliubov Laboratory of Theoretical Physics}
\centerline{\it Joint Institute for Nuclear Research}
\centerline{\it 141980, Dubna, Russia}
\centerline{\it and}
\centerline{\it Institut f\"ur Theoretische Physik }
\centerline{\it Universit\"at Regensburg}
\centerline{\it D-93040 Regensburg, Germany}

\vspace{5mm}

\centerline{\bf D. M\"uller}

\vspace{5mm}

\centerline{\it Institute for Theoretical Physics}
\centerline{\it Center of Theoretical Science}
\centerline{\it Leipzig University}
\centerline{\it 04109 Leipzig, Germany}

\vspace{10mm}

\centerline{\bf Abstract}

\hspace{0.5cm}

Basing on the constraint equalities which arise from the algebra of
the collinear conformal group and the conformal operator product
expansion, we predict the solutions of the leading
order evolution equations for the non-forward distributions, including
transversity, in terms of the conformal moments; next-to-leading order
flavour singlet coefficient functions for the polarized and unpolarized
deeply virtual Compton scattering; as well as the contribution from
renormalon chains to the eigenfunctions of the exclusive non-singlet
evolution kernel.

\end{titlepage}

%%%%%%%%%%%%%%%%%%%%%%%%%%%%%%%%%%%%%%%%%%%%%%%%%%%%%%%%%%%%%%%%%%%%%
\section{Introduction.}
%%%%%%%%%%%%%%%%%%%%%%%%%%%%%%%%%%%%%%%%%%%%%%%%%%%%%%%%%%%%%%%%%%%%%

It is well known that the conformal twist-2 operators possess at
leading order (LO) a diagonal anomalous dimension matrix
$\hat{\gamma}=\{\gamma_{jj'}\}$, which is, in general, triangular
due to Lorentz invariance. One may expect that in the conformal
limit of the theory ($\beta = 0$) the former property will be
fulfilled beyond one-loop order; however, this is not the case
and the conformal operators mix already in the next-to-leading
order (NLO) of perturbation theory. A closer look at the
conformal Ward identities (CWI) in LO for the flavour non-singlet
channel \cite{Mue91,Mue94}, which include both dilatation and special
conformal transformations, shows that the conformal covariance is
destroyed in the minimal subtraction scheme ($\MS$) by an
off-diagonal special conformal anomaly matrix
$\hat{\gamma}^c(l)=\{\gamma^c_{jj'}(l)\}$, which depends on the
spin $l$ carried by the considered twist-2 operator. To restore it to
all orders of perturbation theory one is forced to perform a finite
renormalization of the conformal operators leading to the
conformal subtraction (CS) scheme. Assuming a non-trivial
fixed point $g^*$, i.e. that $\beta (g^*)$ is zero, this
restoration is assured in any order of the perturbation theory by
conformal constraints which arise from the algebra of the collinear
conformal group:
\begin{eqnarray}
\label{conf-constr-KD-1}
\left[ D , K_- \right]_- &=& iK_-
\hspace{2.7cm} \Longrightarrow \hspace{1cm} \left[ \hat a (l)
+ \hat \gamma ^c(l), \hat \gamma \right] = 0, \\
\label{conf-constr-KD-2}
\left[ P_+ , K_- \right]_- &=&
2i( D + M_{-+} )\hspace{1cm} \Longrightarrow \hspace{1cm}
\hat \gamma ^c (l+1) - \hat \gamma ^c (l) = - 2 \hat \gamma ,
\end{eqnarray}
where $a_{jj'}(l)= a(j,l) \cdot \delta_{jj'}$ and $a(j,l)
\equiv 2(j-l)(j+l+3)$. For the case of $\phi^3_{(6)}$ theory,
this constraints can be extended without any assumptions to
nonvanishing $\beta$-function \cite{Mue91}:
\begin{eqnarray}
\label{conf-constr-KD-full}
&&\left[\hat{a}(l)+\hat{\gamma}^c(l)
+2{\beta\over g}\hat{b}(l)
,\hat{\gamma}\right]=0,
\end{eqnarray}
where the $\hat{b}$ matrix is defined as\footnote{ We use the
common definition of the Kronecker symbol
$\delta_{jj'}$ and the
following definition of the (discrete) step-function: $\theta_{jj'} =
\left\{ \begin{array}{c} 1, \ j-j' \geq 0 \\ 0, \ j-j'<0 \end{array}
\right.$. }
\begin{eqnarray}
\label{def-b}
b_{jj'}(l)= \theta_{jj'}
\left\{ 2(l+j'+3)\delta_{jj'} - [1 + (-1)^{ j - j'}] (2j' + 3)
\right\} .
\end{eqnarray}
Hence, the contribution to the off-diagonal matrix elements of the
anomalous dimension matrix due to $\beta$-function
is predicted too. Under the conjecture that only gauge invariant
composite operators contribute to the constraints for the gauge
independent anomalous dimension matrix of the conformal operators,
which turns out to be quite natural, formally the same equation
can also be derived in Abelian gauge field theory for the flavour
non-singlet channel \cite{Mue94} as well as for the quark-gluon
singlet sector studied in the present paper. The knowledge of
the conformal anomaly matrix allows to reconstruct the off-diagonal
part of the anomalous dimension matrix for the corresponding channels.

Few remarks are in order at this point. Although the above mentioned
commutator constraints are derived, for the reasons of technical
simplification, in the theory with $U(1)$ symmetry group,
nevertheless, they are valid in NLO approximation for $QQ$- and
$QG$-channels we have considered, also in the non-Abelian theory.
This is true since in the conformal limit, the symmetry breaking
parts of the kernels do not contain the Casimir operator $C_A$ of
the adjoint representation of $SU(3)$. Moreover, even in the
theory with non-vanishing $\beta$ the correct results can be
reconstructed by substituting the QED $\beta$-function by the
corresponding QCD value \cite{Mue94} and, therefore, this treatment
is sufficient to derive the reliable NLO predictions in QCD. Thus,
some of the results of this paper can be regarded as a first step
towards the complete derivation of the $\alpha_s$-corrections to the
exclusive flavour-singlet evolution kernels which are presently
completely unknown. Using the same technique, the purely gluonic case
will be clarified in the nearest future \cite{BelMul98}. The knowledge
of all other channels will allow one to check the result with a
construction of the ($N=1$) SUSY Ward identities.

In this paper we mainly concern the predictions arising from the conformal
algebra for the deeply virtual Compton scattering (DVCS) process, namely,
the solution of the evolution equations, which govern the $Q^2$-dependence
of the corresponding structure functions, and the NLO corrections to all
singlet coefficient functions in the $\MS$ renormalization scheme. The
final section is devoted to the issue of the renormalon chains contribution
to the Efremov-Radyushkin-Brodsky-Lepage (ER-BL) kernel which can be
incorporated in a straightforward way in our approach.

%%%%%%%%%%%%%%%%%%%%%%%%%%%%%%%%%%%%%%%%%%%%%%%%%%%%%%%%%%%%%%%%%%%%%
\section{Evolution of twist-2 conformal moments at LO.}
%%%%%%%%%%%%%%%%%%%%%%%%%%%%%%%%%%%%%%%%%%%%%%%%%%%%%%%%%%%%%%%%%%%%%

In the past \cite{muel88} and more recently \cite{ji97,rad96,RadPRD56}
different authors introduced non-forward distributions, which
describe the non-perturbative inputs in certain processes such as
DVCS and exclusive electroproduction of mesons. In the consequent
discussion we accept the following definition of the non-forward
distribution functions \cite{rad96,RadPRD56}
\begin{equation}
\label{definition}
{^i{\cal O}^{\mit\Gamma}} (\lambda, \mu)
= \langle h' | \phi_i^* (\mu n) {\mit\Gamma}
\Phi \left[ \mu n, \lambda n \right]
\phi_i (\lambda n) |h \rangle
= \int d x e^{ i \mu (x - \zeta) - i \lambda x }\
{^i{\cal O}^{\mit\Gamma}} (x, \zeta),
\end{equation}
in terms of the light-cone Fourier transformation. Here $i = Q,G$ runs
over the parton species, ${\mit\Gamma}$ corresponds to different Dirac
or Lorentz structures, depending on the spin of the constituents
involved and $\Phi$ is a path ordered exponential. The momentum
fractions of incoming $x$ and outgoing $x - \zeta$ partons are the
Fourier conjugated variables of the light-cone positions $\lambda$
and $\mu$. Parameter $\zeta$ is a skewedness of the distribution
defined as a $+$-component of the $t$-channel momentum $\Delta_+ =
\zeta$ (see next section for discussion of the kinematics of the
DVCS process).

The scale dependence of the light-ray operators,
standing on the left hand side of Eq.\ (\ref{definition}), is
governed by the renormalization group equation which is of the
generic form ($\bar y \equiv 1 - y$)
\begin{eqnarray}
\mu^2\frac{d}{d\mu^2}{\cal O}^{\mit\Gamma} (\lambda , \mu) &=&
\frac{\alpha_s}{2 \pi} \int_{0}^{1} dy \int_{0}^{\bar y} dz
{\cal K}^{\mit\Gamma} (y, z)
{\cal O}^{\mit\Gamma} (\lambda \bar z + \mu z , \mu \bar y + \lambda y).
\end{eqnarray}
The evolution kernels for the parity-even ($V$) and parity-odd ($A$)
chiral-even light-ray operators are known in the literature
\cite{ji97,rad96,RadPRD56,kern,BB88}. Here we supplement this list by the
kernel for the non-forward transversity ($T$) distribution \cite{CFS96}:
\begin{equation}
{\cal K}^T (y, z) = C_F
\left\{ \delta (y) \left[ \frac{1}{z} \right]_+
+ \delta (z) \left[ \frac{1}{y} \right]_+
- \delta (y) - \delta (z) + \frac{3}{2} \delta (y)\delta (z)
\right\}.
\end{equation}
By exploiting the conversion formula derived in Ref. \cite{BM97}
we can easily write the corresponding evolution equation in the
momentum space
\begin{eqnarray}
\label{MOMEE}
\mu^2\frac{d}{d\mu^2}{\cal O}^{\mit\Gamma} (x, \zeta) &=&
- \frac{\alpha_s}{2 \pi} \int d x ' K^{\mit\Gamma} (x , x ', \zeta)
{\cal O}^{\mit\Gamma} (x ', \zeta),
\end{eqnarray}
with\footnote{The explicit expressions as well as integral
and algebraic properties of the $\Theta$-functions introduced
in the main text:
$\Theta^{m}_{i_1 i_2 ... i_n}
(x_1,x_2,...,x_n)=\int_{-\infty}^{\infty}\frac{d\alpha}{2\pi i}
\alpha^m \prod_{k=1}^{n}\left(\alpha x_k -1 +i0 \right)^{-i_k}$
can be found in Ref. \cite{BM97}.}
\begin{eqnarray}
\label{gen-ker-KT}
K^T (x , x ', \zeta)
= C_F \left\{\left[ \frac{x}{x - x'}
\Theta_{11}^0 (x , x - x ') + \frac{x - \zeta}{x - x'}
\Theta_{11}^0 (x - \zeta, x - x ') \right]_+
+ \frac{1}{2} \delta (x ' - x) \right\},
\end{eqnarray}
where $\Theta^0_{11} (x, x') = (\theta(x) - \theta(x'))/(x - x')$. The
corresponding ER-BL kernel ($\zeta=1$)
\begin{eqnarray}
V^T ( x, x' ) \equiv - K^T (x , x', 1)
= C_F
\left\{
\left[
\theta( x' - x ) \frac{x}{x'} \frac{1}{x' - x}
+ \theta(x - x') \frac{\bar x}{\bar x'}\frac{1}{x - x'} \right]_+
+ \frac{1}{2} \delta (x - x')\right\},
\end{eqnarray}
and the DGLAP kernel ($\zeta=0$ and $z=x/x'$)
\begin{eqnarray}
P^T (z) \equiv - K^T (z x' , x', 0)
= C_F
\left\{
2 \left[ \frac{z}{\bar z} \right]_+
+
\frac{1}{2} \delta (\bar z)
\right\}
\end{eqnarray}
follow immediately from the generalized one (\ref{gen-ker-KT}).

In order to find the solution of the evolution equations of the type
given in Eq.\ (\ref{MOMEE}) we are looking for the conformal spin expansion
in terms of reduced non-forward expectation values of conformal twist-2
operators which are related to the Gegenbauer moments of the
corresponding distributions, namely, for quark function we have
\begin{eqnarray}
\zeta^j
\int dx C_j^\frac{3}{2}\left( 2\frac{x}{\zeta} - 1 \right)
{^Q\!{\cal O}^{\mit\Gamma}} (x, \zeta) =
\langle h'| {^Q\!{\cal O}}^{\mit\Gamma}_{jj} |h \rangle .
\end{eqnarray}
$C^\nu_j$ are the Gegenbauer polynomials \cite{BE53_2} with the
normalization fixed below in Eq.\ (\ref{normalization}). The gluon
case differs only in the change of the index of the orthogonal
polynomial by one unity and shifting $j$ to $j - 1$. Then, the
complete sets of conformal twist-2 operators for the flavour
singlet chiral-even un- and polarized as well as chiral-odd
channels read:
\begin{equation}
\label{treeCO}
\left\{\!\!\!
\begin{array}{c}
{^Q\!{\cal O}^V} \\
{^Q\!{\cal O}^A} \\
{^Q\!{\cal O}^T}
\end{array}
\!\!\!\right\}_{jl}
\!=
\bar{\psi} (i \partial_+)^l\!
\left\{\!\!\!
\begin{array}{c}
\gamma_+ \\
\gamma_+ \gamma_5 \\
\sigma_{+\perp} \gamma_5
\end{array}
\!\!\!\right\}
\!C^{\frac{3}{2}}_j\!
\left( \frac{\stackrel{\leftrightarrow}{D}_+}{\partial_+} \right)
\!\psi, \
\left\{\!\!\!
\begin{array}{c}
{^G\!{\cal O}^V} \\
{^G\!{\cal O}^A} \\
{^G\!{\cal O}^T}
\end{array}
\!\!\!\right\}_{jl}
\!=
G_{+ \mu} (i \partial_+)^{l-1}\!
\left\{\!\!\!
\begin{array}{c}
g_{\mu\nu} \\
\epsilon_{\mu\nu+-} \\
0
\end{array}
\!\!\!\right\}
\!C^{\frac{5}{2}}_{j - 1}\!
\left( \frac{\stackrel{\leftrightarrow}{D}_+}{\partial_+} \right)
\!G_{+\nu}.
\end{equation}
Here $\partial \!= \stackrel{\rightarrow}{\partial}\!\!
+\!\! \stackrel{\leftarrow}{\partial}$ and $\stackrel{\leftrightarrow}{D}
= \stackrel{\rightarrow}{\partial}\!\! -\!\! \stackrel{\leftarrow}{\partial}
\!\!- 2i{\rm g}T^aB^a$ and we suppress the flavour and colour indices
for simplicity. The operators we have introduced above mix under the
renormalization and their anomalous dimension matrix
\begin{eqnarray}
\hat{\bf \gamma}=
\left({
{^{QQ}\!\hat{\gamma}}\ {^{QG}\!\hat{\gamma}}\atop
{^{GQ}\!\hat{\gamma}}\ {^{GG}\!\hat{\gamma}}
}\right),
\end{eqnarray}
possesses the square blocks ${^{ik}\!\hat{\gamma}}$ which are
triangular matrices. Since the special conformal anomaly
does not enter in the conformal constraint (\ref{conf-constr-KD-1}) at LO,
the latter takes the simplified form:
\begin{eqnarray}
\label{conf-constr-KD-Sing}
\left[\left({\hat{a}(l)\quad 0\atop\;\;\; 0\quad \hat{a}(l)}\right),
\left({
{^{QQ}\!\hat{\gamma}}\ {^{QG}\!\hat{\gamma}}\atop
{^{GQ}\!\hat{\gamma}}\ {^{GG}\!\hat{\gamma}}
}\right)
\right]=0.
\end{eqnarray}
The matrix $\hat{a}(l)$ is diagonal, then it follows from the
above equality that all entries in $\hat{\bf \gamma}$ are diagonal
too and coincide with the moments of the DGLAP evolution
kernels\footnote{For the $QG$ and $GQ$ sectors this is true up
to some common factors (see Ref. \cite{Cha80} and Eq.\
(\ref{QGmixing}) below).}. Therefore, we can easily write the
solution of the renormalization group equation (\ref{MOMEE})
in terms of the conformal moments, for instance, for the quark non-forward
distribution in the flavour non-singlet channel:
\begin{equation}
\label{solution}
{^{\rm NS}\!{\cal O}^{\mit\Gamma}} (x, \zeta)
= \sum_{j = 0}^{\infty} \frac{1}{N_j (\frac{3}{2}) \zeta^{j + 1}}
\frac{x}{\zeta} \left( 1 - \frac{x}{\zeta} \right)
C_j^{\frac{3}{2}} \left( 2 \frac{x}{\zeta} - 1 \right)
\left( \frac{\alpha_s(Q)}{\alpha_s(Q_0)} \right)^{-
{^{\rm NS}\gamma^{\mit\Gamma}_j}/\beta_0}
\left. \langle h' |
{^{\rm NS}\!{\cal O}^{\mit\Gamma}}_{jj}
| h \rangle \right|_{Q_0^2},
\end{equation}
with $\beta_0 = \frac{2}{3}N_f - \frac{11}{3}C_A$ and the normalization
factor
\begin{equation}
\label{normalization}
N_j (\nu) = 2^{ - 4 \nu + 1 }
\frac{ \Gamma^2 (\frac{1}{2}) \Gamma ( 2 \nu + j )}
{( \nu + j ) \Gamma^2 (\nu) \Gamma (j + 1)}.
\end{equation}
The quark--gluon mixing problem in the singlet channel can be
easily solved by straightforward diagonalization in the same way as
in the case of the deep inelastic scattering (DIS). Note that Eq.\
(\ref{solution}) is valid only for the distributions with a support
$0\le x\le \zeta$ \cite{RadPRD56}.

%%%%%%%%%%%%%%%%%%%%%%%%%%%%%%%%%%%%%%%%%%%%%%%%%%%%%%%%%%%%%%%%%%%%%
\section{NLO coefficient functions for DVCS.}
%%%%%%%%%%%%%%%%%%%%%%%%%%%%%%%%%%%%%%%%%%%%%%%%%%%%%%%%%%%%%%%%%%%%%

In the previous section we have dealt so far with the LO analysis and have
seen that the tree conformal operators (\ref{treeCO}) diagonalize the
anomalous dimension matrix. As we have mentioned in the Introduction, this
property breaks down in the NLO in the $\MS$ scheme even in the theory with
$\beta = 0$. However, the conformal constraints (\ref{conf-constr-KD-1}) and
(\ref{conf-constr-KD-2}) ensure the existence of the CS scheme, in which the
conformal covariance is restored in the conformal limit of the theory to any
order of perturbation theory. In this scheme the OPE in terms of the
conformally covariant operators \cite{FGG73} holds true in the interacting
theory and provides powerful restrictions on the corresponding Wilson
coefficients and can be served, for instance, as a tool for the calculation
of the scattering amplitude of two-photon processes in the generalized
Bjorken region. To compare the predictions with the calculations performed
in the $\MS$ scheme, the following transformation for the operators is
necessary
\begin{eqnarray}
\label{transformation}
{\cal O}_{jl}=\sum_{j'=0}^{j} B_{jj'} \widetilde {\cal O}_{j'l}.
\end{eqnarray}
In the CS scheme the operators $\widetilde O_{jl}$ in different
conformal towers do not mix with each other up to terms proportional
to $\beta$. The transformation matrix $B$, determined completely by
the special conformal anomaly:
\begin{eqnarray}
\label{blefdt-1}
\hat{B} = {\hat{1} \over \hat{1}+{\cal J}\hat{\gamma}^c}
= \hat{1}-{\cal J}\hat{\gamma}^c + {\cal J}(\hat{\gamma}^c
{\cal J}\hat{\gamma}^c) - \dots,
\end{eqnarray}
with ${\cal J}\hat{A} = \theta_{jj' + 1} A_{jj'} / a( j, j' )$,
specifies the partial conformal eigenfunctions of the evolution
kernels. This matrix was calculated in Ref. \cite{Mue94} for the
non-singlet parity-odd channel and used for a consistency check
between the available NLO result for the coefficient function of
the $\gamma^*\gamma \to \pi$ transition form factor and the
forward NLO Wilson coefficients for polarized DIS \cite{Mue97}.

Recently, the $\alpha_s$-correction to the non-forward coefficient
function in the singlet channel has been calculated for unpolarized DVCS
\cite{JiOs97}. In this section we evaluate the one-loop coefficient
functions for the polarized $(T^A_{(1)})$ and non-polarized $(T^V_{(1)})$
cases as well:
\begin{eqnarray*}
{^i T}^{\mit\Gamma}( \omega, x, \zeta, Q^2 | \alpha_s )
= {^i T_{(0)}} ( \omega, x ) - \frac{\alpha_s}{2\pi}
{^i T_{(1)}^{\mit\Gamma}}( \omega, x, \zeta, Q^2 )
+ {\cal O} (\alpha_s^2).
\end{eqnarray*}
The predictions for the coefficient function
can be obtained by employing the  conformal OPE from the one side and
the conformal constraints from the other. In NLO the symmetry
breaking, which is proportional to the $\beta$-function, arises only
in the off-diagonal part (in the basis of Gegenbauer polynomials)
of the anomalous dimension matrix $\hat \gamma$, not in the coefficient
function. So, we are able to make the predictions for the NLO
coefficient functions without restrictions.

To overcome difficulties with gauge invariance and Lorentz
decomposition of the hadronic tensor for the DVCS, we deal, in what
follows, only with appropriate projections of the latter:
\begin{equation}
{\cal F}^{\mit\Gamma} ( \omega, \zeta )
=
i {\cal P}^{\mit\Gamma}_{\mu\nu}
\int d^4 z e^{iqz}
\langle h' |
T \left\{ J_\mu (0) J_\nu (z) \right\}
| h \rangle ,
\end{equation}
where ${\cal P}^V_{\mu\nu} = g_{\mu\nu}$ corresponds to unpolarized
scattering and ${\cal P}^{A}_{\mu\nu} = i \epsilon_{\mu\nu Qp}/(Qp)$
to polarized case. Here $1/\omega = -Q^2/2(Qp)$ and $Q = q - \Delta$.

On the one hand the amplitude for DVCS
can be represented in the factorized form as convolution of the
non-forward distribution (\ref{definition}) and the perturbative
coefficient function
\begin{equation}
\label{DVCS}
{\cal F}^{\mit\Gamma} ( \omega, \zeta ) = \int d x
\sum_{i=Q,G} {^i T^{\mit\Gamma}}( \omega, x, \zeta, Q^2 | \alpha_s )
{^i {\cal O}^{\mit\Gamma}} ( x, \zeta ).
\end{equation}
For the hand-bag diagram\footnote{We omit everywhere the
crossed $u$-channel contribution.} we have to LO in the
coupling  ${^Q T_{(0)}} ( \omega, x ) = \omega /
( x \omega - 1 )$.

On the other hand the conformal OPE \cite{FGG73,Mue97} for the
product of two electromagnetic currents sandwiched between the
hadronic states, in the kinematics appropriate to the process in
question, gives the following prediction:
\begin{equation}
\label{COPE}
{\cal F}^{\mit\Gamma} ( \omega, \zeta ) =
\sum_{i} \sum_{j = 0}^{\infty}
{^iC_j^{\mit\Gamma}} (\alpha_s,Q^2/\mu^2) \
\omega^{j + 1}
{_2F_1} \left( \left.
{
1 + j + {^i\gamma^{\mit\Gamma}_j},\
2 + j + {^i\gamma^{\mit\Gamma}_j}
\atop
4 + 2j + 2\, {^i\gamma^{\mit\Gamma}_j}
}
\right| \omega \zeta \right)
\langle h' | {^i \widetilde {\cal O}}^{\mit\Gamma}_{jj} | h \rangle .
\end{equation}
Here ${_2F_1}$ is confluent hypergeometric function \cite{BE53_1},
the index $i$ sums up the eigenvectors
${^i\widetilde {\cal O}^{\mit\Gamma}}_{jj}$ of the singlet channel,
which possess the diagonal anomalous dimensions
${^i\gamma^{\mit\Gamma}_j}$, and ${^iC_j^{\mit\Gamma}}$ is the
Wilson coefficient of the forward scattering. We have the
convention that in the interacting theory the scale dimension is
shifted by the amount $2{^i\gamma^{\mit\Gamma}_j}$ compared to
canonical one: $d_j = d_j^{\rm can} + 2{^i\gamma^{\mit\Gamma}_j}$.
In Eq.\ (\ref{COPE}) $\mu^2$ is a renormalization scale, while we
have put the factorization scale $\mu^2_F$ equal to $Q^2$. Notice
that for practical purposes $\MS$ schemes will be used where the
manifestly conformaly covariant form of the OPE (\ref{COPE}) is hidden.
Fortunately, the product of two electromagnetic currents is a
renormalization group invariant, so that Eq.\ (\ref{DVCS}) computed
in any scheme is equivalent to Eq.\ (\ref{COPE}) derived in the CS
scheme.

Since the $\alpha_s$-corrections in Eq.\ (\ref{DVCS}) come from the NLO
result for the coefficient function and eigenfunctions of the evolution
kernels we are able to predict the former from the combined use of the
conformal OPE (\ref{COPE}) and the special conformal anomaly which
fixes precisely the corrections to the latter. The general conformal
decomposition of the non-forward distributions looks like
\begin{equation}
{^i{\cal O}} (x, \zeta)
= \sum_{k = Q,G} \sum_{j=0}^{\infty}
{^{ik}\phi}_j (x, \zeta | \alpha_s)
\langle h' | {^k{\cal O}}_{jj} | h \rangle ,
\end{equation}
where the partial conformal waves are generalized, beyond one-loop
level, to non-polynomial functions which are the subject of the
constraints. For the quark case, which is required for the
present application, they have the form
\begin{equation}
{^{Qk}\phi}_j (x, \zeta | \alpha_s)
= \left\{
\delta_{Qk} \delta (x - y)
+ \frac{\alpha_s}{2\pi} {^{Qk}{\mit\Phi}} (x, y, \zeta)
\right\}
\otimes
\frac{1}{N_j (\nu) \zeta^{j + 1}}
\frac{y}{\zeta} \left( 1 - \frac{y}{\zeta} \right)
C_j^{\nu} \left( 2 \frac{y}{\zeta} - 1 \right),
\end{equation}
where $\nu = \nu (k)$ and we introduced the shorthand notation
$\otimes \equiv \int dy$. Here ${^{QQ}{\mit\Phi}}$ is known from
Ref. \cite{Mue94}, while ${^{QG}{\mit\Phi}}$ can be found with the
help of the special conformal anomaly matrix in the quark-gluon
channel. Fortunately, in this sector no subtle conformal symmetry
breaking effects occur as it happens in the quark-quark transition
amplitude. Thus, the special conformal anomaly can be obtained
immediately and reads:
\begin{equation}
\label{QGCAM}
{^{QG}\hat\gamma^{(c)}}
= - \frac{\alpha_s}{2 \pi} \hat{b}(l) \ {^{QG}\hat\gamma}.
\end{equation}

Now we have all necessary results to solve the problem in question.
First we outline the procedure for the evaluation of the coefficient
function for the polarized non-forward gluon contribution and then we
just list the results for other cases since the machinery we have
developed is quite general and can be easily extended to the quark
coefficient functions as well.

%%%%%%%%%%%%%%%%%%%%%%%%%%%%%%%%%%%%%%%%%%%%%%%%%%%%%%%%%%%%%%%%%%%%%
\subsection{Polarized case.}
%%%%%%%%%%%%%%%%%%%%%%%%%%%%%%%%%%%%%%%%%%%%%%%%%%%%%%%%%%%%%%%%%%%%%

For the $QG$-sector we find from Eq.\ (\ref{QGCAM}) that the
correction to the eigenfunctions are completely expressed in terms of
the shift operator:
\begin{equation}
\label{Phi}
{^{QG}{\mit\Phi}} (x, y, \zeta)
= ({\cal I} - {\cal D}) S (x, z) \otimes {^{QG}\!K^A} ( z, y, \zeta ),
\end{equation}
which generates the shift of the Gegenbauer polynomials index:
\begin{equation}
S(x, y) \otimes [y \bar y]^{\nu - \frac{1}{2}}
C^\nu_j (2y - 1)
= \left. \frac{d}{d\rho} \right|_{\rho = 0}
[x \bar x]^{\nu - \frac{1}{2} + \rho} C^{\nu + \rho}_j (2x - 1).
\end{equation}
In Eq.\ (\ref{Phi}) ${\cal I}$ is an identity operator and
${\cal D}$ extracts the diagonal part in the expansion of any test
function $\tau (x, y)$ with  respect to $C^\nu_j$, i.e.
\begin{eqnarray*}
{\cal D}\tau (x, y) = \sum_{ j= 0 }^{\infty}
[x \bar x]^{ \nu - \frac{1}{2}}/ N_j (\nu) C^\nu_j(2x - 1)
\tau_{jj} C^\nu_j (2y - 1).
\end{eqnarray*}
The generalized evolution
kernel\footnote{These generalized kernels posses the following
limits: for ER-BL kinematics $x \leq \zeta $: $K(x , x ',
\zeta) = - V (x/\zeta , x'/\zeta)/\zeta$, for froward DGLAP
evolution $\zeta = 0$: $K(x , x ', 0) = - P(x/x')$.}
\begin{equation}
{^{QG}\!K^A} (x , x ', \zeta)
= 2 N_f T_F \Theta_{112}^1 ( x , x - \zeta, x - x ' )
\end{equation}
can be diagonalized with the following identity
\begin{equation}
\label{QGmixing}
\zeta
\int dy C_{j}^{\frac{3}{2}} \left( 2 \frac{y}{\zeta} - 1 \right)
{^{QG}\!K^A} ( y, x, \zeta )
= {^{QG}\gamma^A_j}
C_{j - 1}^{\frac{5}{2}}
\left( 2 \frac{x}{\zeta} - 1 \right) \ \mbox{with} \
{^{QG}\gamma^A_j} = - \frac{12 N_f T_F}{(j + 1)(j + 2)}.
\end{equation}
The coefficient ${^{QG}\gamma^A_j}$ in front of the Gegenbauer
polynomial coincides with the DGLAP moments of the kernel
${^{QG}\!K^A} ( y, x, 0 )$ up to the common factor $6/j$, which
arises as a result of the conventional definition of the Gegenbauer
polynomials, namely, the coefficient of $x^j$ in $C_{j}^{\frac{3}{2}}
(x)$ is $3/j$ times the coefficient of $x^{j - 1}$ in
$C_{j - 1}^{\frac{5}{2}} (x)$; an additional factor of 2 comes
from the argument of the polynomial.

For the off-diagonal part of the amplitude we get from the
conformal expansion (\ref{COPE}) the following expression
\begin{eqnarray}
\label{nondiagonal}
&&({\cal I} - {\cal D})
\left[
{^Q T_{(0)}} ( \omega, y ) \otimes S ( y, z )
-
{^Q T_{(0)}} ( \omega, z ) \ln ( 1 - z \omega )
\right]
\otimes
{^{QG}\!K^A} ( z, x, \zeta )
\otimes
{^G {\cal O}^A} ( x, \zeta ) \\
&& \hspace{0.5cm} = - 2 \sum_{j = 0}^{\infty}
{^{QG}\gamma^A_j}
B( j + 1 , j + 2 ) \omega^{j + 1}
\left. \frac{d}{d\rho} \right|_{\rho = 0}
{_2F_1} \left( \left.
{
1 + j + \rho,\
2 + j + \rho
\atop
4 + 2j + 2 \rho
}
\right| \omega \zeta \right)
\langle h' | {^G{\cal O}^A}_{jj} | h \rangle . \nonumber
\end{eqnarray}
Its normalization is related to the LO prediction of  the conformal
OPE for the DVCS scattering:
\begin{equation}
{\cal F}^A_{LO} ( \omega, \zeta )
= - 2
\sum_{j = 0}^{\infty}
B( j + 1 , j + 2 )
\omega^{j + 1}
{_2F_1} \left( \left.
{
1 + j ,\
2 + j
\atop
4 + 2j
}
\right| \omega \zeta \right)
\langle h' | {^Q{\cal O}^A}_{jj} | h \rangle ,
\end{equation}

To fix the diagonal part of the hard scattering amplitude we have
to put ${^i\gamma^{\mit\Gamma}_j} = 0$ in Eq.\ (\ref{COPE}), but
keep the NLO result for the forward coefficient function of the
polarized gluon distribution
$\frac{\alpha_s}{2\pi}{^GC^A_j} (Q^2/\mu^2)$ \cite{PGCF,AR88}.
We can write the result as a convolution
\begin{eqnarray}
\label{diagonal}
&&{^Q T_{(0)}} ( \omega, z ) \otimes F ( z, x, \zeta )
\otimes {^G {\cal O}^A} ( x, \zeta ) \\
&& \hspace{0.5cm} = - 2 \sum_{j = 0}^{\infty}
{^GC^A_j} (Q^2/\mu^2)
B( j + 1 , j + 2 ) \omega^{j + 1}
{_2F_1} \left( \left.
{
1 + j ,\
2 + j
\atop
4 + 2j
}
\right| \omega \zeta \right)
\langle h' | {^G{\cal O}^A}_{jj} | h \rangle , \nonumber
\end{eqnarray}
with the following explicit expression for the function $F$, which
corresponds to the DIS coefficient function in the $\MS$ scheme
\cite{AR88}:
\begin{eqnarray}
\label{nondiagonal2}
{^GC^A}_j (Q^2/\mu^2)
&=& 2 N_f T_f \frac{6}{j}
\left\{ ( 2x - 1 )\ln \frac{Q^2\bar x}{\mu^2 x} + 3 - 4x \right\}
\otimes x^j \\
&\Rightarrow& F( z, x, \zeta )
= - {^{QG}\!K^A} (z , x , \zeta) \ln \frac{Q^2}{\mu^2}
+ \left[ 1  - {\cal D} \ln (1 - z \omega) \right]
{^{QG}\!K^A} (z , x , \zeta). \nonumber
\end{eqnarray}

By adding the results of Eqs.\ (\ref{nondiagonal}) and
(\ref{nondiagonal2}) together and subtracting the solution
(\ref{Phi}) of the evolution equation for $\beta = 0$, we finally
come to the coefficient function in the $\MS$ scheme:
\begin{equation}
{^G T_{(1)}^A}( \omega, x, \zeta, Q^2 ) =
{^Q T_{(0)}} ( \omega, y ) \otimes
\biggl\{
{^{QG}\!K^A} ( y, x, \zeta )\, \ln \frac{Q^2}{\mu^2}
+ \left[ \ln \left( 1 - y \omega \right) - 1 \right]
{^{QG}\!K^A} ( y, x, \zeta )
\biggr\}.
\end{equation}

As an outcome of the same but a little bit more involved analysis
for the quark scattering amplitude we get:
\begin{eqnarray}
\label{unpolQCF}
{^Q T_{(1)}^A}( \omega, x, \zeta, Q^2 ) \!\!\!&=&\!\!\!
{^Q T_{(0)}} ( \omega, y ) \otimes
\biggl\{
{^{QQ}\!K^A} ( y, x, \zeta )\,
\ln \frac{Q^2}{\mu^2}
- \frac{3}{2} \left[ {^{Q}\!K^b} ( y, x, \zeta ) \right]_+
+ \frac{3}{2} \delta (y - x) \nonumber \\
&+& \ln \left( 1 - y \omega \right)
{^{QQ}\!K^A} ( y, x, \zeta )
- \left[ {^{Q}\!G} ( y, x, \zeta ) \right]_+
\biggr\} .
\end{eqnarray}
Here
\begin{eqnarray}
{^{QQ}\!K^A} (x , x ', \zeta)
\!\!\!&=&\!\!\! K^T (x , x ', \zeta)
+ C_F \, \Theta_{111}^0 ( x , x - \zeta, x - x ' ), \\
{^{Q}\!G} ( x, x', \zeta )
\!\!\!&=&\!\!\! \frac{x' - \zeta}{x - x'}
\ln \left( \frac{x' - x}{x' - \zeta} \right)
\Theta_{11}^0 (x - \zeta, x - x ')
+
\frac{x'}{x - x'}
\ln \left( \frac{x' - x}{x'} \right)
\Theta_{11}^0 (x, x - x '), \\
{^{Q}\!K^b} ( x, x', \zeta )
\!\!\!&=&\!\!\! \frac{x - \zeta}{x - x'}
\Theta_{11}^0 (x - \zeta, x - x ')
+
\frac{x}{x - x'}
\Theta_{11}^0 (x, x - x '),
\end{eqnarray}
and a new kernel ${^{Q}\!G} ( x, x', \zeta )$ comes from the expression
for ${^{QQ}{\mit\Phi}}$, which is of the form for $\beta = 0$:
\begin{equation}
{^{QQ}{\mit\Phi}} (x, y, \zeta)
= ({\cal I} - {\cal D})
\left\{
S (x, z) \otimes {^{QQ}\!K^A} ( z, y, \zeta )
- \left[ {^{Q}\!G} ( y, x, \zeta ) \right]_+
\right\}.
\end{equation}
In the limit $\omega = 1$ and $\zeta = 1$ we obtain from Eq.
(\ref{unpolQCF}) the results of Refs. \cite{Cha81,Bra83,Rad86}
for the pion transition form factor $F_{\pi\gamma\gamma}$.

%%%%%%%%%%%%%%%%%%%%%%%%%%%%%%%%%%%%%%%%%%%%%%%%%%%%%%%%%%%%%%%%%%%%%
\subsection{Unpolarized case.}
%%%%%%%%%%%%%%%%%%%%%%%%%%%%%%%%%%%%%%%%%%%%%%%%%%%%%%%%%%%%%%%%%%%%%

Since the $QQ$-evolution kernels are the same in the polarized and
non-polarized cases, the only difference will arise from the forward
coefficient functions \cite{ZVN94}. Straightforward calculation
gives for the quark as well as for gluon non-polarized coefficient
functions:
\begin{eqnarray}
{^Q T_{(1)}^V}( \omega, x, \zeta, Q^2 ) \!\!\!&=&\!\!\!
{^Q T_{(0)}} ( \omega, y ) \otimes
\biggl\{
{^{QQ}\!K^V} ( y, x, \zeta )\,
\ln \frac{Q^2}{\mu^2}
- \frac{5}{2} \left[ {^{Q}\!K^b} ( y, x, \zeta ) \right]_+
+ \delta (y - x) \nonumber \\
&+& \left[
\ln \left( 1 - y \omega \right) - 1
\right]
{^{QQ}\!K^V} ( y, x, \zeta )
- \left[ {^{Q}\!G} ( y, x, \zeta ) \right]_+
\biggr\} , \\
{^G T_{(1)}^V}( \omega, x, \zeta, Q^2 ) \!\!\!&=&\!\!\!
{^Q T_{(0)}} ( \omega, y ) \otimes
\biggl\{
{^{QG}\!K^V} ( y, x, \zeta )\,
\ln \frac{Q^2}{\mu^2}
- \frac{1}{2}
\left[
{^{QG}\!K^V} ( y, x, \zeta ) + {^{QG}\!K^A} ( y, x, \zeta )
\right]
\nonumber \\
&+& \ln \left( 1 - y \omega \right) {^{QG}\!K^V} ( y, x, \zeta )
\biggr\} ,
\end{eqnarray}
the momentum space evolution kernels ${^{Q(Q,G)}\!K^V}$ read
\begin{eqnarray}
{^{QQ}\!K^V} (x , x ', \zeta) &=& {^{QQ}\!K^A} (x , x ', \zeta),\\
{^{QG}\!K^V} (x , x ', \zeta)
&=& {^{QG}\!K^A} (x , x ', \zeta)
- 4 N_f T_F \frac{( x - x' )}{x'( x' - \zeta )}
\Theta_{111}^0 ( x , x - \zeta, x - x ' ). \nonumber
\end{eqnarray}
These expressions coincide with the ones obtained in Ref. \cite{JiOs97}
provided we have made some obvious redefinition of their conventions.
Taking the limit intrinsic to the exclusive kinematics we obtain the
well know result \cite{Cha81,Bra83} for the radiative corrections to
the coefficient functions of the parity-even vector meson transition
form factor.

%%%%%%%%%%%%%%%%%%%%%%%%%%%%%%%%%%%%%%%%%%%%%%%%%%%%%%%%%%%%%%%%%%%%%
\section{Renormalon chains in the non-singlet evolution kernel.}
%%%%%%%%%%%%%%%%%%%%%%%%%%%%%%%%%%%%%%%%%%%%%%%%%%%%%%%%%%%%%%%%%%%%%

The last issue of this paper that can be studied by the presently developed
methods is the renormalon chains resummation of the ER-BL evolution kernel.
Recently, this problem has been considered in the $\MS$ scheme for the
non-singlet channel \cite{Mik97,GosKiv97} (see also the work for the DGLAP
splitting kernels \cite{Gra97,Mankiewicz97}). It has been found that the 
resulting series possesses a nonzero convergent radius, so that the infrared 
renormalons are absent in the kernels. From our point of view this result 
can be achieved by a partial resummation of conformal anomalies. The 
$\hat{B}$-matrix defined in Eq.\ (\ref{blefdt-1}) diagonalizes the anomalous 
dimension matrix for the case $\beta=0$. From the conformal constraints 
(\ref{conf-constr-KD-full}) we know that in the $\MS$ scheme its off-diagonal 
part is induced by $\hat{\gamma}^c + 2\frac{\beta}{g}\hat{b}$. Therefore, 
the operator 
\begin{eqnarray}
\hat{B} =
{\hat{1} \over \hat{1} + {\cal J}
\left(\hat{\gamma}^c + 2 \frac{\beta}{g}\hat{b}\right)}
\end{eqnarray}
diagonalize the anomalous dimension matrix\footnote{Since the
$\hat{B}$-matrix depends on the running coupling, the diagonalization
of the anomalous dimension matrix provide the solution of the
renormalization group equation only in the fixed coupling regime.}.
This formula can be understood as a resummation prescription for the
conformal anomalies that induce the off-diagonal part. Knowing this
matrix one can construct the eigenfunctions
$\phi_l (x | \alpha_s) \equiv {^{QQ}\phi}_l (x, 1 | \alpha_s)$
of the ER-BL kernel \cite{Mue94}
\begin{equation}
\phi_l(x | \alpha_s)=
\sum_{j=l}^\infty \frac{ x \bar x }{N_j(\frac{3}{2})}
C_j^{\frac{3}{2}}( 2 x - 1 ) B_{jl}(\alpha_s).
\end{equation}
In Ref.\ \cite{Mue94} a differential equation for the eigenfunctions
was derived in the conformal limit. Repeating the steps, it is easy
to restore the neglected $\beta$ term. Employing the relation (see
Appendix in \cite{Mue97})
\begin{eqnarray}
a(j, l) B_{jl}(\alpha_s)
= - \left\{\left(\hat{\gamma}^c(l)
+ 2 \frac{\beta}{g}\hat{b}(l)\right)\hat{B}\right\}_{jl}
\end{eqnarray}
and the eigenvalue equation for the Gegenbauer polynomials
\cite{BE53_2} we are able to derive an integro-differential
equation for the eigenfunctions in the full theory, valid for the $\MS$
scheme:
\begin{equation}
\label{MSequation}
x \bar x \frac{d^2}{dx^2} \phi_l(x | \alpha_s)
+ ( l + 1 )( l + 2 ) \phi_l(x | \alpha_s)=
\frac{1}{2} \int_0^1 dy\,
\left(
\gamma_l^c(x,y,\alpha_s) + 2 \frac{\beta}{g} b_l(x,y)
\right)
\phi_l(y | \alpha_s),
\end{equation}
where the kernels can be obtained from their non-forward analogues
introduced in the preceding sections and read
\begin{eqnarray}
\gamma_l^c(x,y,\alpha_s) &=& \frac{\alpha_s}{2\pi}
\left\{
2 \int_0^1 dz\, b_l(x,z) \left[ {^{QQ}V}(z,y) \right]_+
+ [g(x,y)]_+
\right\} +O(\alpha_s^2), \\
b_l(x,y)&=&  2 \left( l + 1 + \frac{ 2x - 1}{2} \frac{d}{dx} \right)
\delta( x - y ), \\
g(x,y)&=& C_F \theta( y - x ) \frac{\ln(1 - \frac{x}{y})}{ y - x }
+ \left\{{ x \to \bar x \atop y \to \bar y }\right\}, \\
{^{QQ}V} (x,y)&=&
C_F \theta( y - x )
\frac{x}{y} \left( 1 + \frac{1}{ y - x } \right)
+ \left\{ {x \to \bar x \atop y \to \bar y } \right\}.
\end{eqnarray}
The non-perturbative solution of Eq. (\ref{MSequation}) is unknown.
However, in the large $N_f$ limit only the term proportional to
$\beta$ survives. Taking into account also the non-leading $N_f$
terms included in the $\beta$-function, the equation (\ref{MSequation})
can be reduced to
\begin{eqnarray}
&&\hspace{-1cm}x \bar x \frac{d^2}{dx^2}\
\phi_l(x | \alpha_s)
+ ( l + 1 )( l + 2 ) \phi_l(x | \alpha_s) \\
&&\hspace{3cm} = \frac{\beta}{g} \int_0^1 dy\,
\left\{
2 \left( l + 1 + \frac{ 2 x - 1 }{2} \frac{d}{dx} \right)\delta(x-y) +
O(\alpha_s)
\right\}\
\phi_l (y | \alpha_s).
\nonumber
\end{eqnarray}
Taking the known LO contribution on the right hand side yields a
differential equation which has the solution:
\begin{eqnarray}
\label{eigfunc-beta}
\phi_j(x | \alpha_s)
\sim
\left[ x \bar x \right]^{\frac{3}{2} - \frac{\beta}{g}}
C_j^{\frac{3}{2}-\frac{\beta}{g}}( 2 x - 1 ).
\end{eqnarray}
This result coincide in the large $N_f$ limit with the resummation of
the vacuum polarization bubbles in Refs.\ \cite{Mik97,GosKiv97}
and verifies the `naive non-abelianization' hypothesis employed there
\cite{GosKiv97}. Moreover, it is very interesting to see that this
hypothesis can be extended in our case to the {\em whole} $\beta$-function.
Unfortunately, the eigenfunctions (\ref{eigfunc-beta}) have no practical
importance for the solution of the evolution equation. Obviously, in
this treatment the special anomaly matrix contribution is completely
discarded, however, it was demonstrated by numerical studies in Ref.
\cite{Mue95} that a considerable cancellation occurs between the former
and the symmetry breaking term due to the $\beta$-function for the real
case $N_f=3$.

%%%%%%%%%%%%%%%%%%%%%%%%%%%%%%%%%%%%%%%%%%%%%%%%%%%%%%%%%%%%%%%%%%%%%
\section{Summary.}
%%%%%%%%%%%%%%%%%%%%%%%%%%%%%%%%%%%%%%%%%%%%%%%%%%%%%%%%%%%%%%%%%%%%%

In the present study we employ the powerful restrictions from the
conformal algebra on the amplitudes in the massless theory. In the
leading order of perturbation theory it enables us to give the
solution of the evolution equations for the off-forward distribution
functions in terms of the conformal partial wave expansion. Beyond
the leading order the conformal symmetry breaking term induced by
nonvanishing $\beta$-function does not appear in the coefficient
function so that we can rely on the hypothetical conformal limit of
the theory in order to make the predictions for the two-photon
processes in the generalized Bjorken kinematics. To this end the
knowledge of the forward Wilson coefficients in the OPE and the
diagonal anomalous dimensions of the corresponding operators which
were determined earlier from the study of deep inelastic scattering
is required .

The development of the present method is far from being finished
since up to now all the results were derived in the theory with
$U(1)$ symmetry group. The extension of the conformal constraints
to purely Yang-Mills theory \cite{BelMul98} will allow to get much
deeper insight into the conformal symmetry breaking by the
non-abelian fields.

\vspace{0.5cm}

We wish to thank B. Geyer, S.V. Mikhailov, D. Robaschik, A. Sch\"afer
for valuable discussions. D.M. would like to thank the Theory Group
at the Joint Institute for Nuclear Research  for the kind hospitality
during his stay, where the initial part of this work has been done.
A.B. was supported by the Russian Foundation for Fundamental
Research, grant N 96-02-17631 and Deutsche Physikalische Gesellschaft.
D.M. was financially supported by the Deutsche Forschungsgemeinschaft
(DFG).

\end{document}